\documentclass[conference]{IEEEtran}
\IEEEoverridecommandlockouts
\usepackage{cite}
\usepackage{amsmath,amssymb,amsfonts}
\usepackage{algorithmic}
\usepackage{graphicx}
\usepackage{textcomp}
\usepackage{xcolor}
\usepackage{braket}
\usepackage{tikz}
\usetikzlibrary{quantikz2}
\usepackage{amsfonts,amssymb,amsmath}

\usepackage{environ}
\usepackage{mathtools}
\usepackage{adjustbox}
\usepackage{multirow}

\NewEnviron{gather+}[1][1]{%
  \begin{equation*}
  \scalebox{#1}{$\begin{gathered}\BODY\end{gathered}$}
  \end{equation*}
}

\def\BibTeX{{\rm B\kern-.05em{\sc i\kern-.025em b}\kern-.08em
    T\kern-.1667em\lower.7ex\hbox{E}\kern-.125emX}}
\begin{document}

\title{Towards Distributed Quantum Error
Correction for Distributed Quantum
Computing\\
}

\author{\IEEEauthorblockN{Shahram Babaie}
\IEEEauthorblockA{\textit{Department of Computer Science and Engineering} \\
\textit{University at Buffalo}\\
Buffalo, USA \\
shahramb@buffalo.edu}
\and
\IEEEauthorblockN{Chunming Qiao}
\IEEEauthorblockA{\textit{Department of Computer Science and Engineering} \\
\textit{University at Buffalo}\\
Buffalo, USA \\
qiao@buffalo.edu}
\and
}

\maketitle

\begin{abstract}
Quantum computing as a promising technology can utilize stochastic solutions instead of deterministic approaches for complicated scenarios for which classical computing is inefficient, provided that both the concerns of the error-prone nature of qubits and the limitation of the number of qubits are addressed carefully. In order to address both concerns, a new qubit-based Distributed Quantum Error Correction (DQEC) architecture is proposed in which three physical qubits residing on three Quantum Processing Units (QPU) are used to form a logical qubit. This paper illustrates how three QPUs collaboratively generate a joint quantum state in which single bit-flip and phase-flip errors can be properly resolved. By reducing the number of qubits required to form a logical qubit in the proposed architecture, each QPU with its limited number of physical qubits can accommodate more logical qubits than when it has to devote its three physical qubits for each logical qubit. The functional correctness of the proposed architecture is evaluated through the Qiskit tool and stabilizer generators. Moreover, the fidelity of input and output quantum states, the complexity of the proposed designs, and the dependency between error probability and correctness of the proposed architecture are analyzed to prove its effectiveness.
\end{abstract}

\begin{IEEEkeywords}
Distributed quantum computing, Error correction, Complexity, Fidelity
\end{IEEEkeywords}

\section{Introduction}
Quantum computing as a nascent technology leverages quantum mechanics to actualize probabilistic tools for complex computations. There is a broad consensus among the academic and industrial communities regarding quantum computing, as this technology promises to solve complex problems, develop new materials, accelerate complicated optimization problems, come up with new ideas in communication areas, and handle intricate dynamics in financial markets. This technology is based on peculiar phenomena of quantum physics such as interference, superposition, and entanglement that cannot be explained through classical physics laws such as Maxwell’s equations and Newton’s laws \cite{10064036}. 

The strength and capabilities of a quantum computer directly depend on the number of its qubits, which also determines the dimension of Hilbert space or state space of that system and its computational capabilities. Although the number of qubits has so far been growing even faster than Moore’s law, from IBM's 27-qubit chip in 2019 to the 1,121-qubit chip in 2023, there are several hurdles to increasing the number of qubits and achieving large-scale quantum computers \cite{9528073}. In addition, available quantum systems are in the Noisy Intermediate-Scale Quantum \textit{‘‘NISQ’’} era because they are subject to errors and there are several disruptive factors such as faulty preparation, faulty measurement, defective gates, imperfect qubits, and disorders caused by interaction with the environment. Moreover, coherent and incoherent errors drive quantum states into a random and featureless state, which is an unforgivable destructive factor in stored quantum information and information preservation elements. It should be noted that quantum systems tend to interact with the environment and become entangled or coupled with it, causing a decaying excited state and dephasing. Furthermore, in contrast to classical computers that are almost flawless, quantum computers are prone to failure due to defective faulty gates and imperfect qubits, where the transistor error rate is $p_c \approx 10^{-27}$ while the qubit error rate is $p_q \approx 10^{-3}$ \cite{AutomaticImplementationandEvaluation}. 

Besides the above-mentioned factors, there are other destructive factors such as qubit drop and qubit leakage, which disrupt the operation of quantum systems. Together, these factors will preclude the supremacy of quantum technology in the absence of effective mitigation strategies. In fact, all quantum algorithms such as Grover’s algorithm, Shor’s algorithm, and Simon’s algorithm assume quantum gates and qubits are perfect, therefore, such rigorous theoretical foundations, such as database searching and factoring are only effective with flawless components \cite{444}. Quantum errors are also major threats to quantum networks and quantum Internet which are growing applications of quantum technology. It should be noted that the errors are cumulative and increase exponentially with the increase of circuit depth. Therefore, Quantum Error Correction (QEC) techniques play an indispensable role in tackling defective factors and achieving fault-tolerant systems, deeper quantum circuits, and large-scale quantum computing \cite{10497274}.

Quantum Error Correction Codes (QECCs) are a combination of information theory, quantum mechanics, and classical theory of error-correcting codes. There are several challenges regarding quantum error correction techniques that need to be addressed carefully. In general, information redundancy is an effective strategy for error detection and correction, however, the no-cloning theorem limits utilizing the classical concept of information redundancy in the quantum domain \cite{10483460}. Moreover, the correctness assessment of a state or a register can only be accomplished through observation, but measurement collapses the waveform of quantum states, rendering them useless for further operations. Furthermore, in contrast to the bit-flip error correction in classical systems, an effective quantum error correction technique should tackle bit-flip, phase-flip, and decoherence errors, which can be easily introduced by faint disruptive factors such as feeble magnetic fields and stray microwave pulses. A prevalent approach in qubit-based QEC is to use multiple physical qubits to form logical qubits that are more fault-tolerant than physical qubits. However, this strategy further limits the computing scale of a QPU, since the number of qubits available on a QPU is already quite limited \cite{10461424}.

Recently, Distributed Quantum Computing (DQC) has received a lot of attention as a new paradigm for increasing quantum computing power through the collaboration of multiple inter-connected small-to-moderate scale Quantum Processing Units (QPU) instead of an ordinarily monolithic quantum unit \cite{caleffi2022distributed,9488850}. At the time of writing, Condor, IBM’s largest processor, has 1,121 qubits, while it is expected that the next product will support distributed structure, including 4158 qubits with three Kookaburra processors. Nevertheless, it has been shown that tens or hundreds of thousands or more qubits are needed for advanced applications. For example, it is estimated that Shor’s algorithm needs around 20 million physical qubits to factor a 2048-bit number. A distributed quantum computing system with thousands or more QPUs interconnected via a quantum data network \cite{9812920,9912163} seems to be an effective solution to approach this number of qubits in the coming years. Within such a distributed quantum computing system, a large number of qubits needed for a quantum circuit can be split into many QPUs \cite{1111}. This concept of DQC, along with the need for QEC, has motivated this research on distributed quantum error correction
\cite{QuantumDataNetworking,2222}.

\subsection{Paper Contributions}
In this paper, a new Distributed Quantum Error Correction (DQEC) technique is presented that can correct single bit-flip and phase-flip errors of quantum-based systems. The contributions of the proposed architecture can be summarized as follows.

\begin{itemize}
 \renewcommand{\labelitemi}{\scriptsize$\blacksquare$}
 \item Distribution of the error correction algorithm into three different quantum processing units. 
 \item Correction of single bit-flip and phase-flip errors in distributed architecture and analysis of its correctness through stabilizer generators.
 \item Assessment of the proposed qubit-based DQEC with a universal error set in Qiskit.
 \item Analysis of fidelity between quantum information source and quantum information sink to evaluate the correctness of the proposed DQEC.
 \item Determination of error probability threshold to clarify error tolerance of the proposed architecture. 
\end{itemize}

\subsection{Paper Organization}
The remainder of this paper is organized as follows: Section 2 provides preliminaries and quantum channels. Section 3 provides details of the proposed qubit-based architecture to correct bit-flip and phase-flip in the collaboration of three QPUs. The simulation results and evaluation of the proposed approach with a universal error set are presented in section 4. Section 5 provides a brief review of related works. Finally, section 6 concludes the paper and provides some guidelines for future research.

\section{Preliminaries and Error Channels}
\subsection{Quantum States}
The basic difference between classical computing and quantum computing is related to the differences between bit and qubit. Bit is the basic unit of information in classical computing, which can be either 0 or 1 at any particular instant \cite{10247656}. By contrast, qubit, a basic unit of quantum information, is in a superposition state that is a linear combination of orthonormal vectors $\ket{0}$ and $\ket{1}$ in $\mathbb{C}^2$ before measurement or observation. Therefore, a quantum state can be represented in \textit{ket} notation or Dirac notation as $\ket{\psi} = \alpha \ket{0}+\beta\ket{1}$, where $\alpha,\beta \in \mathbb{C}$ and are called probability amplitudes. Measurement collapses the wave function of a qubit irreversibly through projection to $\ket{0}$ or $\ket{1}$ with the probability of $|\alpha|^2$ and $|\beta|^2$, respectively, according to Born’s rule. Moreover, the probability amplitude of a valid quantum state satisfies the second axiom of probability theory, $|\alpha|^2 + |\beta|^2=1$ \cite{10171549}.

In general, a qubit can be represented through a superposition state vector in 2-dimensional Hilbert space, while an \textit{N-qubit} composite system is associated with a $2^N$-dimensional Hilbert space. A 2-qubit composite system can be either in an entangled state or product state. Entanglement explains how two subatomic particles can be so tightly correlated that they no longer act independently even though their distance is billions of light-years \cite{AsynchronousEntanglementProvisioning}. Entangled particles, as integral parts of DQC, cannot be separated into the tensor product of component systems. For instance, $\alpha \ket{00} + \beta\ket{11}$ can not be written as the tensor product of two states as $[\alpha_1\ket{0} + \beta_1\ket{1}] \otimes [\alpha_2\ket{0}+\beta_2\ket{1}]$. There are four entangled states, as depicted in Equation \ref{Eq1}, known as Bell states or EPR states that are used in most applications as well as distributed quantum computing \cite{SegmentedEntanglement}.
\begin{equation} \label{Eq1}
\begin{split}
    \ket{\Phi^+} = \frac{\ket{00}+\ket{11}}{\sqrt{2}} \qquad \qquad \ket{\Psi^+} = \frac{\ket{01}+\ket{10}}{\sqrt{2}} \\
    \ket{\Phi^-} = \frac{\ket{00}-\ket{11}}{\sqrt{2}} \qquad \qquad \ket{\Psi^-} = \frac{\ket{01}-\ket{10}}{\sqrt{2}}
\end{split}
\end{equation}

\subsection{Quantum Errors}
In this subsection, the sources and various types of errors in quantum computing are discussed first.  Note that many small errors can accumulate and cause bit-flip or phase-flip errors. Quantum computing is error-prone due to destructive factors arising from a variety of factors, including environmental interactions, imperfect control mechanisms, and the intrinsic fragility of quantum states. It has been shown that effective error mitigation strategies are necessary to achieve fault tolerance, reliable, and dependable quantum computing, in which each type of disruptive factor requires specific logistics. Since all points on the surface of the block sphere are valid, a small rotation around any axis changes a pure state to another pure state, making it challenging to distinguish the original state from the changed state. In particular, there are several potential sources of error in quantum technology as follows. The first step of a quantum algorithm is the preparation step where all qubits are initialized to a specific state such as $\ket{\psi}$. There is no efficient way to ensure the quantum state of qubits after preparation. If the desired state is $\ket{\psi}$, it is perfect if there is evidence that the post-preparation quantum state is a pure state such as $\ket{\psi'}$ even though $\ket{\psi} \neq \ket{\psi'}$, because there is a full knowledge about the quantum state after preparation \cite{8423050}. In reality, a post-preparation quantum state is a distribution of states instead of a pure state, which is a statistical ensemble of pure states that the density operator can describe these mixed states, $\rho = \sum_{i} p_i\ket{\psi_i}\bra{\psi_i}$, where $p_i$ represents the probability of being $i^{th}$ state, $\ket{\psi_i}$, after preparation. The last step of a quantum algorithm is measurement which is accomplished through the interaction with the quantum system, and it has been proven that any type of interaction is fraught with error. Measurement error can be modeled through Positive Operator Value Measure (POVM) and Projection Valued Measure (PVM) \cite{10188776}. 

Qubits in a QPU are non-isolated and intrinsically tend to interact with the environment and entangle with the environment, which is studied through quantum decoherence theory \cite{10361038}. Interaction with the environment can cause to decay of quantum information stored in the system, the decaying excited state, $\ket{1}$, to the ground state, $\ket{0}$. This process is called spontaneous emission and the time required for this change is called relaxation time or longitudinal time $(T_1)$. Moreover, this interaction can cause dephasing which refers to the loss of coherent superposition of a quantum state, and the time required for this change is called dephasing time $(T_2)$. Relaxation errors and dephasing errors altogether are called decoherence errors and their corresponding times are called decoherence times \cite{IntegratingAll-opticalSwitching}. If we consider $\ket{0}_E$ and $\ket{1}_E$ as the basis states of the environment, any interaction of a quantum state with the environment can be modeled as Equation \ref{Eq2}.
\begin{gather} 
        \ket{0}_Q\ket{0}_E \rightarrow \ket{0}_Q\ket{0}_E \notag\\
        \ket{1}_Q\ket{0}_E \rightarrow \sqrt{1-\omega} \ket{1}_Q\ket{0}_E + \sqrt{\omega} \ket{0}_Q\ket{1}_E
        \label{Eq2}
\end{gather}
where $\omega$ is the probability of photon loss rate, which is also called damping probability, and its value at time instant \textit{t} is equal to $\omega = 1-e ^ {\frac{-t}{T_1}}$. In this equation, the environment is initialized to the vacuum state, $\ket{0}_E$. According to this model, the quantum state will be unchanged if it is in the ground state, $\ket{0}_Q$, while it will be changed to the ground state with the probability of $\omega$ and the environment state will be changed from $\ket{0}_E$ to $\ket{1}_E$ if the quantum state is excited. Therefore, the entanglement of a quantum state $\ket{\psi} = \alpha\ket{0}+\beta\ket{1}$  with the environment can be represented as Equation \ref{Eq3}.
\begin{equation} \label{Eq3}
    \ket{\psi}\ket{0}_E \rightarrow (\alpha\ket{0}+\beta \sqrt{1-\omega} \ket{1})\ket{0}_E +\sqrt{\omega}\beta\ket{0}\ket{1}_E
\end{equation}

 In addition, quantum computers suffer from defective faulty gates and imperfect qubits in contrast to classical computers which are almost flawless. Although the error rate of each quantum gate or qubit may be small, a small error in a large number of consecutive gates may change the outcome, called cumulative error in quantum algorithms. Let's suppose there is a quantum circuit consisting of $N$ identity gates arranged in series, the circuit is initiated with $\ket{0}$, and each identity operator has a small error probability as $\epsilon$. The final output is $\ket{\psi}_{out}=\prod_{i}^{N} I_i\ket{0}$ and the expected output is $\ket{0}$, where $I\equiv \sigma_x$. While cumulative of $\epsilon$ error changes the output value to $\ket{\psi}_{out}=\prod_{i}^{N} e^{i\epsilon}\sigma_x\ket{0} = \cos{N\epsilon}\ket{0}+i\sin{N\epsilon}\ket{1}$. Such that the output will be $\ket{0}$ and $\ket{1}$ with the probability of $P(\ket{0})=\cos ^2 {(N\epsilon})\simeq 1-(N\epsilon)^2$ and $P(\ket{1})=\sin ^2 {(N\epsilon})\simeq (N\epsilon)^2$, respectively, instead of being in the pure state $\ket{0}$. Therefore, the probability of error in this single-type-gate quantum circuit is $P_{error} \simeq (N\epsilon)^2$. Consequently, the cumulative feature of error implies that any small rotation on the Bloch sphere in any quantum gate quadratically changes the expected output \cite{AnAsynchronousTransportProtocol}. 

\subsection{Quantum Channel Models}
Quantum errors can be classified into operational errors and loss errors. Measurement error, memory decoherence, and imperfect gates are examples of operational errors, while photon absorption in the fiber and detector inefficiency are examples of loss errors. Moreover, depending on the error effects, quantum errors are classified into coherent errors and incoherent errors. Coherent errors are deterministic unitary operations and can be pictured as incorrect rotations on the Bloch sphere. These errors can be corrected straightforwardly like regular hardware calibration. For example, applying rotation around the \textit{x-axis} by $\pi$ radians, $\textit{X}\ket{0}=R_{\overrightarrow{x}} (\pi) = e ^ {-i\pi\frac{X}{2}}$ on $\ket{0}$ converts it to $\ket{1}$, $e ^ {-i\pi\frac{X}{2}}\ket{0} = \ket{1}$. In other words, the arrow pointing towards the north pole is changed towards the south pole by applying \textit{X-gate}. Any over-rotation or under-rotation, caused by an imperfect \textit{X-gate} error leads to incorrect rotation and wrong resultant state as $\Tilde{X}\ket{0}=e ^ {-i(\pi \pm \epsilon)\frac{X}{2}}\ket{0} \neq \ket{1}$. The resultant state is not an ideal state, while it is a pure state and can be represented by a point on the Bloch sphere. This is a coherent error and can be dispelled through hardware calibration. Incoherent errors, called stochastic errors, can be non-deterministic unitary operations or non-unitary operations, which can be represented through the Pauli channel and relaxation channel, respectively \cite{10477919}.

In classical information theory, a channel as a space-domain element refers to a transmission line that transmits classical information between distinct and independent systems, while in quantum information theory, a channel as a time-domain concept represents the factors of quantum state evolution over time \cite{3333}. In other words, quantum channels represent quantum state changes through transformation over time. Binary Symmetric Channel (BSC) is an analogous concept in classical computing, which represents the probability of correct reception or flipped probability of information at the receiver side. Pauli channel is an effective way to model the quantum errors that correspond to the influence of Pauli operators on quantum states. Pauli gates as Pauli operators $(\sigma_x,\sigma_y,\sigma_z)$ act on a single qubit and represent rotation around the $\textit{x}$, $\textit{y}$, and $\textit{z}$ axes of the Bloch sphere by $\pi$ radians, respectively. Pauli operators and $\sigma_0$ as zeroth Pauli matrix or identity matrix are represented as Equation \ref{Eq4}.

\begin{equation} \label{Eq4}
\resizebox{.91\hsize}{!}{$
\sigma_x= \left(\begin{array}{ccc}
0 & 1\\
1 & 0\end{array}\right)
\sigma_y= \left( \begin{array}{ccc}
0 & -i\\
i & 0\end{array} \right)
\sigma_z= \left( \begin{array}{ccc}
1 & 0\\
0 & -1\end{array} \right)
\sigma_0= \left( \begin{array}{ccc}
1 & 0\\
0 & 1\end{array} \right)
$}
\end{equation}
The Pauli $\sigma_x$, denoted by \textit{X}, is a bit-flip operator that converts $\ket{0}$ to $\ket{1}$ or vice versa, analogous to the NOT gate in classical computing. The general representation of this operator is depicted in Equation \ref{Eq5}.
\begin{equation} \label{Eq5}
    \textit{X}\ket{\psi} = \left( \begin{array}{ccc}
0 & 1\\
1 & 0\end{array} \right)\left( \begin{array}{c}
\alpha\\
\beta\end{array} \right) = \beta\ket{0}+\alpha \ket{1}
\end{equation}
Pauli $\textit{X-channel}$, $\eta_x(\rho)$, can be modeled as Equation \ref{Eq6}, which maps an input pure state to the mixed state in the density matrix representation, where $p_x$ denotes the bit-flip error rate.
\begin{equation} \label{Eq6}
    \eta_x(\rho) = (1-p_x)\rho + p_xX\rho X
\end{equation}
The Pauli $\sigma_z$, denoted by Z, is a phase-flip operator that changes the phase of $\ket{1}$ and does not affect $\ket{0}$. The general representation of this operator is depicted in Equation \ref{Eq7}.
\begin{equation} \label{Eq7}
    \textit{Z}\ket{\psi} = \left( \begin{array}{ccc}
1 & 0\\
0 & -1\end{array} \right)\left( \begin{array}{c}
\alpha\\
\beta\end{array} \right) = \alpha\ket{0}-\beta \ket{1}
\end{equation}    
The bit/phase-flip error can be represented by Pauli $\sigma_y$, denoted by Y, defined as Equation \ref{Eq8}, which is a combination of Pauli $\sigma_z$ and Pauli $\sigma_x$, as $\sigma_y=i\sigma_x\sigma_z$.
\begin{equation}\label{Eq8}
        \textit{Y}\ket{\psi} = \left( \begin{array}{ccc}
0 & -i\\
i & 0\end{array} \right)\left( \begin{array}{c}
\alpha\\
\beta\end{array} \right) = -i(\alpha\ket{0}-\beta \ket{1})
\end{equation}
Pauli \textit{Z-channel} and \textit{Y-channel}, $\eta_z(\rho)$ and $\eta_y(\rho)$, can be written similarly to Equation \ref{Eq8}, by substituting $p_z$ and $p_y$ as phase-flip error rate and Y error rate, respectively, and Z and Y as single-bit operators. The Pauli channel $\eta_p(\rho)$ is a combination of these three quantum channels, which maps the input state onto a linear combination of four different operations that can be presented as Equation \ref{Eq9}.
\begin{equation}\label{Eq9}
\resizebox{.91\hsize}{!}{$
\eta_p(\rho) = (1-p_x-p_y-p_z)I\rho I+p_xX\rho X+p_yY\rho Y+p_zZ\rho Z $}
\end{equation}
where \textit{I} represents a 2×2 identity operator, which leaves the quantum state unchanged, and $p_x$,$p_y$, and $p_z$ entirely depend on qubit relaxation time $(T_1)$ and dephasing time $(T_2)$ and can be calculated through Equation \ref{Eq10}.
\begin{equation} \label{Eq10}
    p_x=p_y=\frac{1}{4}(1-e^{-\frac{t}{T_1}}) \qquad p_z=\frac{1}{4}(1+e^{-\frac{t}{T_1}}-2e^{-\frac{t}{T_2}})
\end{equation}
It should be noted that $p_x$ and $p_y$ just depend on relaxation time, while $p_z$ depends on both relaxation time and dephasing time. Although phase-flip occurs more often than bit-flip and bit/phase-flip and most quantum systems behave as asymmetric channels, in a special case, the error rate of three cases is assumed equal and the channel is called the depolarizing channel, denoted by $\eta_{DP}(\rho)$. Therefore, Equation \ref{Eq9} can be rewritten as Equation \ref{Eq11}.
\begin{equation}\label{Eq11}
    \eta_{DP}(\rho) = (1-p)I\rho I+\frac{p}{3}(X\rho X+Y\rho Y+Z\rho Z)    
\end{equation}
The amplitude damping channel is a well-known model that represents the energy loss in a generalized quantum system that maps an input state in a mixed state to a new mixed state as Equation \ref{Eq12}.
\begin{equation} \label{Eq12}
       \eta_{DP}(\rho) = E_0\rho E_0\textsuperscript{\textdagger}+E_1\rho E_1\textsuperscript{\textdagger}
\end{equation}
where $E_0$ and $E_1$ represent error operators and are called \textit{Kraus} operators whose matrix representations are $ \resizebox{.25\hsize}{!}{$ E_0=\begin{bmatrix}
1 & 0 \\
0 & \sqrt{1-\omega} 
\end{bmatrix} $}$ and $ \resizebox{.25\hsize}{!}{$E_1=\begin{bmatrix}
0 & \sqrt{\omega} \\
0 & 0 \end{bmatrix}$}$. Phase damping is another channel model to represent the dephasing effect of environmental decoherence that leads to the loss of quantum information without loss of energy. The \textit{Kraus} operators of this channel are $ \resizebox{.25\hsize}{!}{$E_0=\begin{bmatrix}
1 & 0 \\
0 & \sqrt{1-\gamma} 
\end{bmatrix}$}$ and $ \resizebox{.25\hsize}{!}{$E_1=\begin{bmatrix}
0 & 0 \\
0 & \sqrt{\gamma} \end{bmatrix}$}$, where $\gamma$ represents the scattering probability of a photon in the realistic systems, which is equal to $\gamma = 1-e^{\frac{t}{T_1}-\frac{2t}{T_2}}$ at time instant \textit{t} \cite{9741787}.

\section{The proposed DQEC Architecture}
\begin{figure*}
	\centering
	\includegraphics[width=.7\textwidth]{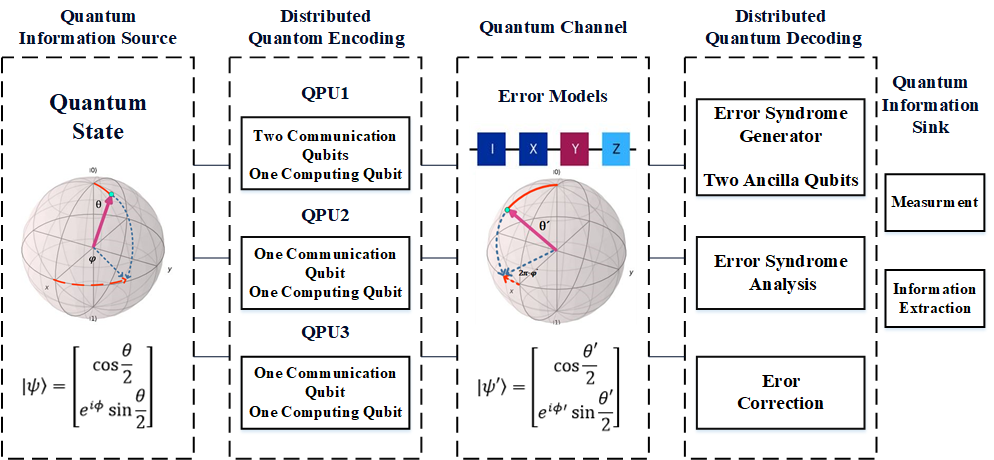}
	\caption{General overview of the proposed architecture.}
	\label{Fig1}
\end{figure*}

In general, destructive factors of quantum systems can lead to either error qubits or erasure qubits in quantum information. The error refers to circumstances where the receiver knows there is an error in information but does not know which one, while erasure refers to a circumstance where the receiver knows which qubit is missing but does not know its value. In a monolithic quantum system, three physical qubits including one computing qubit and two ancilla qubits are needed to form a logical qubit \cite{8423050}. Consequently, only about $\frac{N}{3}$ logical qubits can be formed with $N$ physical qubits residing on a QPU. Due to the limitation of the number of physical qubits residing on a QPU, a transition from monolithic quantum computing to distributed procedures seems to be inevitable in order to increase the number of available logical qubits. This is because regardless of the advance in technology, there is a threshold for the number of physical qubits residing on a QPU to be cost-effective. Therefore, when the number of qubits of a QPU, $N$, reaches this threshold, increasing $N$ further becomes a lot more difficult than increasing the number of QPUs, each with $N$ qubits. The proposed architecture is an endeavor in this direction that focuses on the distribution of bit-flip and phase-flip error correction processes on distinct QPUs. Furthermore, the proposed architecture endeavors to diminish the number of physical qubits required per QPU, which is the paramount limitation of quantum computing in the available versions. The general structure of the proposed architecture is illustrated in Figure \ref{Fig1}.

The quantum information source generates a quantum state as a linear superposition of computational basis, $\ket{\psi}=\alpha\ket{0} +\beta\ket{1}$, which should be delivered correctly to the quantum information sink. As shown in Figure \ref{Fig1}, the proposed architecture consists of three modules, i.e., distributed encoding, quantum channel or error channel, and distributed decoding. The distributed encoder creates a joint state as a logical qubit of the input quantum state $(\ket{\psi})$ resided on QPU\#1 and utilizes two computing qubits resided on QPU\#2 and QPU\#3. The Quantum channel receives $\ket{\psi}$ as a valid quantum state, applies $\ket{\psi}_e$ as an error state to a valid state through utilizing unitary operation $(U_e )$, and forwards it to the decoding segment and finally, three QPUs collaboratively perform distributed decoding, the details of which will be discussed further.

\subsection{Distributed Quantum Encoding}
\begin{figure}[hbt!]
\begin{adjustbox}{width=\columnwidth,center}
\begin{quantikz}
       \quad \lstick{A} & \ctrl{3} &   \ctrl{5}    &&&&&&   & \measure{\sigma_z} && \measure{\sigma_z} &\rstick[3]{out}\\  
        \quad \lstick{B} &&    &&& \targ{}  &&&&&&& \\
        \quad \lstick{C} &&   &&&&  \targ{} &&&&&& \\
       \lstick[2]{$\ket{\Phi^+}$} \push{A1} \quad & \targ{} & & \gate{Z}   \\
       \push{B1} \quad &&& \targ{} \wire[u][1]{q} &&  \ctrl{-3} &&& \gate{H} & \gate{Z} \wire[u][4]{q}\\
       \lstick[2]{$\ket{\Phi^+}$} \push{A2}\quad && \targ{} &&  \gate{Z}            \\
       \push{C1}\quad  &&&& \targ{} \wire[u][1]{q} &&\ctrl{-4} && \gate{H} &&& \gate{Z} \wire[u][6]{q}
\end{quantikz}
\end{adjustbox}
\caption{Distributed encoding segment of the proposed architecture.}
\label{Fig2}
\end{figure}
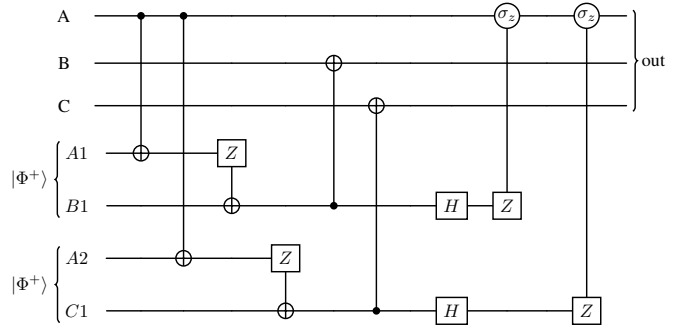

Encoding refers to the convention of data-word (\textit{k}) to code-word (\textit{n}) to enhance data protection through adding information redundancy. Due to the use of information redundancy (\textit{c}), the code-word can be retrieved even if some qubits are erroneously flipped. Encoding mechanisms, which are entirely distinguished by coding strategies, are different in terms of the amount of information redundancy, the position of redundant bits in code-words, and the calculation process of information redundancy. Repetition of the data-word to generate the code-word is the simplest linear encoding mechanism. Let’s assume $\textit{M}$ indicates the number of repetitions, therefore, Hamming distance (\textit{d}) of the repetition code is equal to $\textit{M}$ and the code rate (\textit{R}) is equal to $\frac{1}{M}$. Due to the no-cloning theorem, which limits the cloning of an arbitrary quantum state, repetition-based techniques are impossible in quantum computing. Therefore, in the proposed architecture, the entanglement-based technique is applied to achieve information redundancy where three different QPUs collaborate to accomplish the encoding section. The quantum state $\ket{\psi}=\alpha\ket{0}+\beta\ket{1}$ as a linear superposition of computational basis is the original state that should be protected against bit-flip and phase-flip, which is placed in QPU\#1. Moreover, two ancillary qubits initialized to $\ket{0}$ are selected from QPU\#2 and QPU\#3. Figure \ref{Fig2} illustrates the logical collaboration of three QPUs to create a joint state of three qubits.

The proposed encoding segment is implemented through three QPUs belonging to Alice (\textit{A}), Bob (\textit{B}), and Charlie (\textit{C}), for example, to generate a joint state of three qubits. There are two entangled pairs between Alice and Bob $(A_1$ and $B_1)$, and between Alice and Charlie $(A_2$ and $C_1)$, which are denoted by $\ket{\Phi^+}$ in Figure \ref{Fig2}. The quantum state between these states can be represented as Equation \ref{Eq13}.
\begin{equation}\label{Eq13}
\resizebox{.89\hsize}{!}{$
    \ket{\phi}_{A_1B_1A_2C_1}=\frac{1}{\sqrt{2}}(\ket{00}+\ket{11})_{A_1B_1}\otimes\frac{1}{\sqrt{2}}(\ket{00}+\ket{11})_{A_2C_1}=\ket{\Phi^+}_{A_1B_1}\otimes\ket{\Phi^+}_{A_2C_1}
    $}
\end{equation}
where $A$,$A_1$, and $A_2$ belong to QPU\#1, $B$ and $B_1$ belong to QPU\#2, and $C$ and $C_1$ belong to QPU\#3. It should be noted that $A$,$B$, and $C$ are computing qubits while $A_1$,$A_2$,$B_1$ and $C_1$ are communication qubits. The initial state of three computing qubits, $\ket{\psi}_{ABC}$, can be represented as Equation \ref{Eq14}.
\begin{equation}\label{Eq14}
\resizebox{.89\hsize}{!}{$
    \ket{\psi}_{ABC}=(\alpha\ket{0}+\beta\ket{1})_A\otimes\ket{0}_B\otimes\ket{0}_C=(\alpha\ket{000}+\beta{\ket{100}})_{ABC}
    $}
\end{equation}
where the initial state of $A$ is $\alpha\ket{0}_A+\beta\ket{1}_A$, which is necessarily arbitrary, and the initial states of ancilla qubits are $\ket{0}_B$ and $\ket{0}_C$. This quantum state is contributed with entangled states to generate the total system state as Equation \ref{Eq15}.
\begin{multline}
\label{Eq15}
\ket{\psi}= \ket{\psi}_{ABC}\otimes\ket{\phi}_{A_1B_1A_2C_1}=(\alpha\ket{000}+\beta\ket{100})_{ABC} \\
\otimes\frac{1}{\sqrt{2}}(\ket{00}+\ket{11})_{A_1B_1}\otimes\frac{1}{\sqrt{2}}(\ket{00}+\ket{11})_{A_2C_1}
\end{multline}

The encoding process is as follows.

\textbf{Step 1:} There are two local CNOT gates as $CNOT(A,A_1)$ and $CNOT(A,A_2)$, which generate a general state as in Equation \ref{Eq16}.

\begin{multline}
\label{Eq16}
CNOT(A, A_1), CNOT(A, A_2):\ket{\psi}=
\frac{1}{2}[(\alpha\ket{000}\otimes\\
(\ket{00}+
\ket{11}))\otimes(\ket{00}+\ket{11})+\beta\ket{100}\otimes(\ket{10}+\ket{01})\otimes\\(\ket{10}+\ket{01})]_{ABCA_1B_1A_2C_1}
=\alpha\ket{000}_{ABC}\otimes \ket{\Phi^+}_{A_1B_1}\otimes\\\ket{\Phi^+}_{A_2C_1}+
\beta\ket{100}_{ABC} \otimes\ket{\Psi^+}_{A_1B_1}\otimes\ket{\Psi^+}_{A_2C_1}
\otimes\\
\frac{1}{\sqrt{2}}(\ket{00}+\ket{11})_{A_1B_1}\otimes\frac{1}{\sqrt{2}}(\ket{00}+\ket{11})_{A_2C_1}
\end{multline}

\textbf{Step 2:} QPU\#1 applies computational basis single-qubit measurement to $A_1$ and $A_2$ and notifies their results to QPU\#2 and QPU\#3, respectively. Then QPU\#2 and QPU\#3 apply controlled-NOT$(Pauli-X)$ operation to $B_1$ and $C_1$ if the $Z-basis$ measurement result is $\ket{1}$. The quantum state after step 2 can be represented as Equation \ref{Eq17}.
\begin{equation}\label{Eq17}
\resizebox{.89\hsize}{!}{$
   \ket{\psi}=\frac{1}{2}[\alpha\ket{000}_{ABC}\otimes\ket{00}_{B1C1}+\beta\ket{100}_{ABC}\otimes\ket{11}_{B1C1}]$} 
\end{equation}
Therefore, $A_1$ and $A_2$ are removed from the quantum state, denoting $A_1$ and $A_2$ are terminated at this stage. Qubits $B_1$ and $C_1$ contain the necessary information about the qubit $A$.

\textbf{Step 3:} QPU\#2 and QPU\#3 apply local CNOT as $CNOT(B_1,B)$ and $CNOT(C_1,C)$ to their local qubits as in Equation \ref{Eq18}, which describes the transmission of an arbitrary quantum state of QPU\#1 to both QPU\#2 and QPU\#3.
\begin{equation}\label{Eq18}
\resizebox{.89\hsize}{!}{$
    \ket{\psi}=\frac{1}{2}[\alpha\ket{000}_{ABC}\otimes\ket{00}_{B1C1}+\beta\ket{111}_{ABC}\otimes\ket{11}_{B1C1}]$}   
\end{equation}
\textbf{Step 4:} At this step, the quantum states of $B_1$ and $C_1$ should be removed from the quantum state. This can be accomplished by applying the Hadamard gate, computational basis measurement, and a controlled Pauli-Z operation to qubit $A$. The quantum state after applying the Hadamard gate to $B_1$ and $C_1$ is in the form of Equation \ref{Eq19}.
\begin{equation}\label{Eq19}
\resizebox{.89\hsize}{!}{$
\begin{gathered}
\ket{\psi}=\frac{1}{2}[\alpha\ket{000}_{ABC}\otimes\frac{1}{2}(\ket{00}_{B_1C_1}+\ket{01}_{B_1C_1}+\ket{10}_{B_1C_1}
+\ket{11}_{B_1C_1})\\
+\beta\ket{111}_{ABC}\otimes\frac{1}{2}(\ket{00}_{B_1C_1}-\ket{01}_{B_1C_1}-\ket{10}_{B_1C_1}+\ket{11}_{B_1C_1})]
\end{gathered}
$}
\end{equation}

The general state after applying the computational basis measurement to $B_1$ and $C_1$ and applying the controlled-Z to qubit $A$ is in the form of Equation \ref{Eq20}.
\begin{equation}\label{Eq20}
\begin{split}
\ket{\psi}=&\frac{1}{4}[4\alpha\ket{000}_{ABC}+4\beta\ket{111}_{ABC}]=\\
&\alpha\ket{000}_{ABC}+\beta\ket{111}_{ABC}
\end{split}
\end{equation}
The generated state is a joint state of three qubits residing on three different QPUs. Therefore, the data-word is one qubit and the code-word consists of three qubits, which makes it clear that the code rate is equal to $R=\frac{k}{n}=\frac{1}{3}=0.33$.

 In general, quantum errors can be modeled through quantum channels. The Pauli \textit{X-gate} and $R_{\overrightarrow{n}}(\alpha)$, as a rotation around the \textit{n-axis} by $\alpha$ radians are applied to model the quantum errors. Let’s assume there are $m$ quantum channels between each pair of QPUs, therefore, there are 2m+1 different combinations for occurring single bit-flip, single phase-flip, and error-free cases, because of the possibility of occurring either bit-flip or phase-flip on each channel and one case as an error-free case. For example, $I\otimes Z \otimes I$ and $I\otimes I \otimes X$ represent phase-flip on the second channel and bit-flip on the third channel, respectively. It should be noted that consideration of various values for $\alpha$ within $[0,2\pi)$ provides numerous phase-shifting cases as $R_{\overrightarrow{n}}(\alpha)=e^{-i\frac{\alpha}{2}\overrightarrow{n}.\sigma}$ that should be analyzed in the correctness assessment of the proposed architecture.

\subsection{Distributed Quantum Decoding}
Decoding refers to the analysis of the received code-word $(n)$, error syndrome calculation, extraction of data-word $(k)$ from code-word, and detection or correction of possible errors. In general, error control techniques can be classified into two groups, i.e., ARQ and FEC. Forward Error correction (FEC) techniques involve detecting and correcting a subset of possible errors through the attached information redundancy, while Automatic Repeat reQuest techniques (ARQ) can only detect the errors and issue a re-transmission request. Due to the no-cloning theorem and temporal evolution of quantum states, quantum components do not have access to the previous states and the idea of re-transmission in the quantum region would be impractical. Therefore, effective quantum decoders should try to correct errors instead of planning a complementary mechanism. In classical techniques, the exact position of the error bit is necessary for the error correction process, while quantum error correction techniques can perform error correction without having to specify the exact location of the error bit. In classical techniques, the received information can be observed/measured and the detected error bit is flipped back easily for error correction. As shown,  bit-flip and phase-flip errors can be corrected in QEC  without having to know which physical qubit is flipped. Nevertheless, additional effort can be made to generate error syndrome in order to facilitate further analysis such as the identification of error-prone qubits or a certain failure pattern to aid in the operation, administration, and maintenance (OA\&M) of the system.

\begin{figure*}
    \centering
    \resizebox{0.75\textwidth}{!}{%
 \begin{quantikz}
\lstick{A}  & \ctrl{2} &   \ctrl{1}    &&&&&&   & \measure{\sigma_z} & \measure{\sigma_z}&&&& \targ{} &&&&\rstick[3]{out}  \\
\push {\ket{\phi^+_{2}}} \push{A2} \; &&\targ{} &&&& \gate{Z} &&&&&&& \targ{}&\ctrl{1} && \gate{X} &\\
\qquad \qquad \push {\ket{\Phi^+_{1}}} \push{A1} \;  &\targ{} &&& \gate{Z} &&&&&&&& \targ{} &&\ctrl{-2} & \gate{X} &&\\
\qquad \qquad \push{B} \;  &&&&& \targ{} &&&&&&  \ctrl{1} & &&&\measure{\sigma_z} \wire[u][1]{q}\\
\qquad \qquad \push {\ket{\Phi^+_{1}}} \push{B1}  \; &&&&\targ{} \wire[u][2]{q} & \ctrl{-1} &&& \gate{H} & \gate{Z} \wire[u][4]{q}  && \targ{} & \gate{Z} \wire[u][2]{q}\\ \lstick{C} &&&&&&& \targ{}  &&&& \ctrl{1} & &&&&\measure{\sigma_z} \wire[u][4]{q} \\
\push {\ket{\phi^+_{2}}} \push{C1} \; &&&&& & \targ{} \wire[u][5]{q}  & \ctrl{-1} & \gate{H} && \gate{Z} \wire[u][6]{q} & \targ{} && \gate{Z} \wire[u][5]{q}
\end{quantikz}
}%
\caption{Distributed decoding segment of the proposed architecture.}
\label{Fig3}
\end{figure*}
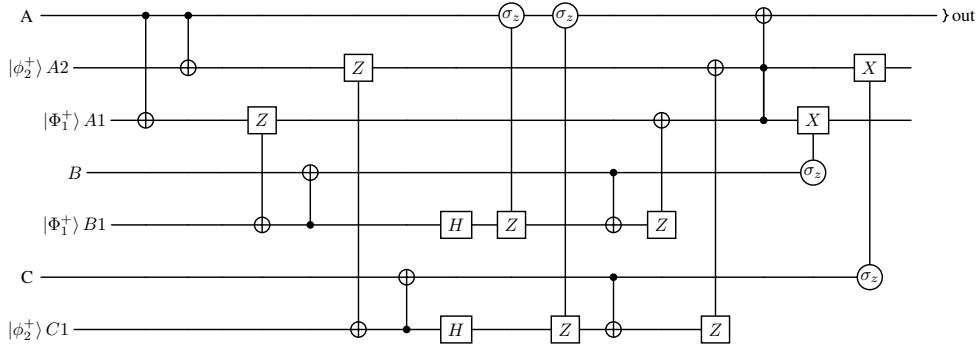

The proposed distributed decoding segment receives a three-qubit joint state that should correct any errors that may occur. Figure \ref{Fig3} illustrates the logical collaboration of three QPUs to correct bit-flip errors in the three-qubit joint state.

 Similar to the encoding segment, three QPUs collaborate for the distributed decoding segment. The decoding segment can be split into two distinct segments, i.e., error correction segment and quantum information extraction segment. The error correction segment is responsible for correcting the received state $\ket{\psi}_e$ and covert it to a valid state still includes information redundancy, while the extraction segment tries to eliminate information redundancy and extract the arbitrary quantum state that is utilized at the initialization step of the encoding segment. There are two entangled pairs between $A_1$ and $B_1 (\ket{\Phi^+_1})$, and between $A_2$ and $C_1(\ket{\Phi^+_2})$ to realize the collaboration of three QPUs. The simplest bell state, $\ket{\phi}_{A_1B_1A_2C_1}=\ket{\Phi^+}_{A_1B_1}\otimes\ket{\Phi^+}_{A_2C_1}$ is utilized in the proposed design.The distributed encoding segment produced the $\ket{\psi}=\alpha\ket{000}_{ABC}+\beta\ket{111}_{ABC}$ in the last step. The bit-flip error can occur independently with equal probability for each qubit. Therefore, the input state of the decoding segment will be one of the following four possible states. i.e., $\ket{\psi}_e=\alpha\ket{000}_{ABC}+\beta\ket{111}_{ABC}$, $\ket{\psi}_e=\alpha\ket{001}_{ABC}+\beta\ket{110}_{ABC}$, $\ket{\psi}_e=\alpha\ket{010}_{ABC}+\beta\ket{101}_{ABC}$, or $\ket{\psi}_e=\alpha\ket{100}_{ABC}+\beta\ket{011}_{ABC}$ with equal probabilities, which respectively represent error-free, first qubit error, second qubit error, and third qubit error. Therefore, the general quantum state of the input’s decoding segment can be written as Equation \ref{Eq21}.
\begin{equation}\label{Eq21}
\resizebox{.89\hsize}{!}{$
\begin{gathered}
    \ket{\psi}_e=[\frac{1}{4}(\alpha\ket{000}_{ABC}+\beta\ket{111}_{ABC})+\frac{1}{4}(\alpha\ket{001}_{ABC}+\beta\ket{110}_{ABC})+\frac{1}{4}(\alpha\ket{010}_{ABC}\\+\beta\ket{101}_{ABC})
    +\frac{1}{4}(\alpha\ket{100}_{ABC}+\beta\ket{011}_{ABC})]_{ABC}\otimes\ket{\Phi^+}_{A_1B_1}\otimes\ket{\Phi^+}_{A_2C_1}
\end{gathered}
$}
\end{equation}
The decoding process is as follows.

\textbf{Step 1:} Two local CNOT gates are utilized between $A$ and $A_1$ and between $A$ and $A_2$ within QPU\#1, which generate a quantum state as Equation \ref{Eq22}.
\begin{equation}\label{Eq22}
\begin{split}
&\ket{\psi}_e=\frac{1}{4}[\alpha\ket{000}_{ABC}\otimes\ket{\Phi^+}_{A_1B_1}\otimes\ket{\Phi^+}_{A_2C_1}+\beta\ket{111}_{ABC}\\&
\otimes\ket{\Psi^+}_{A_1B_1}\otimes\ket{\Psi^+}_{A_2C_1}]
+\frac{1}{4}[\alpha\ket{001}_{ABC}\otimes\ket{\Phi^+}_{A_1B_1}\otimes\\&
\ket{\Phi^+}_{A_2C_1}+\beta\ket{110}_{ABC}\otimes\ket{\Psi^+}_{A_1B_1}\otimes\ket{\Psi^+}_{A_2C_1}]\\&
+\frac{1}{4}[\alpha\ket{010}_{ABC}\otimes\ket{\Phi^+}_{A_1B_1}\otimes\ket{\Phi^+}_{A_2C_1}+\beta\ket{101}_{ABC}\otimes\\&
\ket{\Psi^+}_{A_1B_1}\otimes\ket{\Psi^+}_{A_2C_1}]
+\frac{1}{4}[\alpha\ket{100}_{ABC}\otimes
\ket{\Phi^+}_{A_1B_1}\otimes\\&
\ket{\Phi^+}_{A_2C_1}+
\beta\ket{011}_{ABC}\otimes\ket{\Psi^+}_{A_1B_1}\otimes\ket{\Psi^+}_{A_2C_1}]
\end{split}
\end{equation}
\textbf{Step 2:} At this step, computational basis single-qubit measurement is applied to $A_1$ within QPU\#1, and the result is notified to QPU\#2, and another computational basis single-qubit measurement is applied to $A_2$ within QPU\#1 and the result is notified to QPU\#3. QPU\#2 and QPU\#3 apply controlled-NOT$(Pauli-X)$ operation to $B_1$ and $C_1$ if their corresponding $Z-basis$ measurement results are $\ket{1}$, otherwise, do nothing. The general quantum state after applying these operations is as Equation \ref{Eq23}.

\begin{equation}\label{Eq23}
\begin{split}
&\ket{\psi}_e=\frac{1}{4}[\alpha\ket{000}_{ABC}\otimes\ket{00}_{B_1C_1}+\beta\ket{111}_{ABC}\otimes\ket{11}_{B_1C_1}]\\&
+\frac{1}{4}[\alpha\ket{001}_{ABC}\otimes\ket{00}_{B_1C_1}
+\beta\ket{110}_{ABC}\otimes\ket{11}_{B_1C_1}]\\&
+\frac{1}{4}[\alpha\ket{010}_{ABC}\otimes\ket{00}_{B_1C_1}+\beta\ket{101}_{ABC}\otimes\ket{11}_{B_1C_1}]\\&
+\frac{1}{4}[\alpha\ket{100}_{ABC}\otimes\ket{11}_{B_1C_1}+\beta\ket{011}_{ABC}\otimes\ket{00}_{B_1C_1}]
\end{split}
\end{equation}

\textbf{Step 3:} QPU\#2 applies local CNOT as $CNOT(B_1,B)$ and QPU\#3 applies local CNOT as $CNOT(C_1,C)$ to their local qubits as in Equation \ref{Eq24}. 
\begin{equation}\label{Eq24}
\begin{split}
&\ket{\psi}_e=\frac{1}{4}[\alpha\ket{000}_{ABC}\otimes\ket{00}_{B_1C_1}+\beta\ket{100}_{ABC}\otimes\ket{11}_{B_1C_1}]\\&
+\frac{1}{4}[\alpha\ket{001}_{ABC}\otimes\ket{00}_{B_1C_1}
+\beta\ket{101}_{ABC}\otimes\ket{11}_{B_1C_1}]\\&
+\frac{1}{4}[\alpha\ket{010}_{ABC}\otimes\ket{00}_{B_1C_1}+\beta\ket{110}_{ABC}
\otimes\ket{11}_{B_1C_1}]\\&
+\frac{1}{4}[\alpha\ket{111}_{ABC}\otimes\ket{11}_{B_1C_1}+\beta\ket{011}_{ABC}\otimes\ket{00}_{B_1C_1}]
\end{split}
\end{equation}
\textbf{Step 4:} QPU\#2 and QPU\#3 apply the Hadamard gate to $B_1$ and $C_1$, followed by a computational basis measurement and applying a controlled Pauli-Z operation to qubit $A$ in QPU\#1 based on the measurement results. After applying these three unitary operations, the general quantum state is in the form of Equation \ref{Eq25}.
\begin{equation}\label{Eq25}
\begin{split}
&\ket{\psi}_e=\frac{1}{4}[(\alpha\ket{000}_{ABC}+\beta\ket{100}_{ABC})+\\&(\alpha\ket{001}_{ABC}+
\beta\ket{101}_{ABC})
+(\alpha\ket{010}_{ABC}+\\&
\beta\ket{110}_{ABC})+
(\alpha\ket{111}_{ABC}+\beta\ket{011}_{ABC})]
\end{split}
\end{equation}
\textbf{Step 5:} The entangled pairs between $A_1$ and $B_1$ and between $A_2$ and $C_1$ have been broken at step 2 after applying computational basis measurement. It is necessary to establish entangled states between QPU\#1 and QPU\#2 and between QPU\#1 and QPU\#3 for subsequent operations. This can be conducted through various techniques, which in this research it is assumed that the EPR pairs are currently available. Therefore, the general state of this step considering the EPR pairs can be written as Equation \ref{Eq26}.
\begin{equation}\label{Eq26}
\begin{split}
&\ket{\psi}_e=[\frac{1}{4}(\alpha\ket{000}_{ABC}+\beta\ket{100}_{ABC})+\frac{1}{4}(\alpha\ket{001}_{ABC}+\\&
\beta\ket{101}_{ABC})+\frac{1}{4}(\alpha\ket{010}_{ABC}
+\beta\ket{110}_{ABC})+\frac{1}{4}\\&
(\alpha\ket{111}_{ABC}+\beta\ket{011}_{ABC})]_{ABC}\otimes\ket{\Phi^+}_{A_1B_1}\otimes\ket{\Phi^+}_{A_2C_1}
\end{split}
\end{equation}
\textbf{Step 6:} QPU\#2 applies a local CNOT as $CNOT(B,B_1)$ and QPU\#3 applies a local CNOT as $CNOT(C,C_1)$ to their local qubits as in Equation \ref{Eq27}.
\begin{equation}\label{Eq27}
\begin{split}
&\ket{\psi}_e=\frac{1}{4}[(\alpha\ket{000}_{ABC}+\beta\ket{111}_{ABC})\otimes\ket{\Phi^+}_{A_1B_1}\\&
\otimes\ket{\Phi^+}_{A_2C_1}+(\alpha\ket{001}_{ABC}+\beta\ket{101}_{ABC})
\otimes\\&\ket{\Phi^+}_{A_1B_1}\otimes
\ket{\Psi^+}_{A_2C_1}+(\alpha\ket{010}_{ABC}+\beta\ket{110}_{ABC})\\&\otimes\ket{\Psi^+}_{A_1B_1}
\otimes
\ket{\Phi^+}_{A_2C_1}+(\alpha\ket{111}_{ABC}+\beta\ket{011}_{ABC})\\&\otimes\ket{\Psi^+}_{A_1B_1}
\otimes\ket{\Psi^+}_{A_2C_1}]
\end{split}
\end{equation}

\textbf{Step 7:} At this step, QPU\#2 and QPU\#3 apply a computational basis single-qubit measurement to $B_1$ and $C_1$, respectively, and notify the results to QPU\#1 to apply controlled-NOT $(Pauli-X)$ operation to $A_1$ and $A_2$ if their corresponding $Z-basis$ measurement results are $\ket{1}$ otherwise, do nothing. The general state after applying these operations is depicted as Equation \ref{Eq28}.
\begin{equation}\label{Eq28}
\begin{split}
\ket{\psi}_e=&\frac{1}{4}[(\alpha\ket{000}_{ABC}+\beta\ket{100}_{ABC})\otimes\ket{00}_{A_1A_2}+\\&
(\alpha\ket{001}_{ABC}+\beta\ket{101}_{ABC})\otimes\ket{01}_{A_1A_2}
+\\&(\alpha\ket{010}_{ABC}
+\beta\ket{110}_{ABC})\otimes\ket{10}_{A_1A_2}+\\&
(\alpha\ket{111}_{ABC}+\beta\ket{011}_{ABC})\otimes\ket{11}_{A_1A_2}]
\end{split}
\end{equation}
\textbf{Step 8:} QPU\#1 applies a local Toffoli gate between its qubits $A_1$ and $A_2$ as control qubits and $A$ as the target qubit. The general quantum state after applying this operation is as Equation \ref{Eq29}. 
\begin{equation}\label{Eq29}
\begin{split}
\ket{\psi}_e=&\frac{1}{4}[(\alpha\ket{000}_{ABC}+\beta\ket{100}_{ABC})\otimes\ket{00}_{A_1A_2}+\\&
(\alpha\ket{001}_{ABC}+\beta\ket{101}_{ABC})\otimes\ket{01}_{A_1A_2}+\\&
(\alpha\ket{010}_{ABC}+\beta\ket{110}_{ABC})\otimes\ket{10}_{A_1A_2}+\\&(\alpha\ket{011}_{ABC}+\beta\ket{111}_{ABC})\otimes\ket{11}_{A_1A_2}]
\end{split}
\end{equation}
\textbf{Step 9:} QPU\#1 applies a Hadamard basis measurement to $A_1$ and $A_2$, and sends their results to QPU\#2 and QPU\#3, respectively. QPU\#2 and QPU\#3 apply the Pauli-Z operation to qubit $B$ and qubit $C$ if the measurement results are $\ket{-}$. After applying these unitary operations, the general quantum state can be represented as Equation \ref{Eq30}. 
\begin{equation}\label{Eq30}
\begin{split}
&\ket{\psi}_e=\frac{1}{4}[(\alpha\ket{000}_{ABC}+\beta\ket{100}_{ABC})+\\&(\alpha\ket{000}_{ABC}
+\beta\ket{100}_{ABC})
+(\alpha\ket{000}_{ABC}+\\&\beta\ket{100}_{ABC})+(\alpha\ket{000}_{ABC}+\beta\ket{100}_{ABC})]
\end{split}
\end{equation}
According to the final quantum state, the output of the distributed decoding segment is always $\alpha\ket{000}_{ABC}+\beta\ket{100}_{ABC}$. In other words, the quantum state of $A$ is equal to $\ket{\psi}_A=\alpha\ket{0}_{A}+\beta\ket{1}_{A}$, which is the same as the input quantum state, and the ancilla qubits are returned to $\ket{0}$, as the initial state. Therefore, the proposed distributed encoding and decoding structures can correct all possible single bit-flip errors.

\subsection{Syndrome Analysis}
\begin{figure}[hbt!]
\begin{adjustbox}{width=\columnwidth,center}
\begin{quantikz}
\lstick[3]{Quantum Channel Outputs} \push{A} \quad && \ctrl{3} && \ctrl{4} &&\\
\push{B} \quad &&& \ctrl{2} &&&\\
\push{C} \quad &&&&& \ctrl{2}&\\
\lstick[2]{Ancilla Qubits} \push{\ket{0}}\quad && \targ{}&\targ{} &&& \meter{}&\rstick[2]{$\ket{S_e}$}\\ 
\push{\ket{0}} \quad &&&& \targ{} & \targ{} & \meter{} & 
\end{quantikz}
\end{adjustbox}
\caption{Error syndrome generator.}
\label{Fig4}
\end{figure}
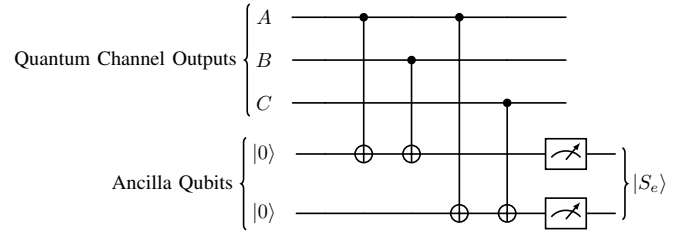

The proposed distributed decoder can automatically detect and correct all single bit-flip errors but there is no error pattern to clarify the position of the erroneous qubit. Figure \ref{Fig4} illustrates the error syndrome generator that can be utilized before the decoding segment to produce the error pattern if needed. Statistical analysis of error syndrome provides more knowledge about identifying error sources and manufacturing new products with more calibrated hardware components. According to the generated error syndrome, the decoding segment distinguishes the exact erroneous qubit as follows. Since the measurement collapses the qubit wave function and projects the quantum state on basis states, the location of the errors must be determined without measurement, based on which two auxiliary qubits are generated and measured if needed.

This circuit has two ancilla qubits initialized with $\ket{0}$ corresponding to the no error case. Two-bit syndrome $\ket{S_e}$ generator is based on a two-by-two comparison of qubits. $\ket{S_0}$ is generated through the comparison of the first and second qubits and $\ket{S_1}$ is generated through the comparison of the first and third qubits. After applying CNOT gates if $\ket{S_0}$ is still equal to zero it means that the first and second qubits are the same, but otherwise, they are different. Likewise, if $\ket{S_1}$ is still equal to zero it means that the first and third qubits are the same, but otherwise, they are different. The erroneous qubit can be identified by analyzing the error syndrome $\ket{S_1S_0}$. The $\ket{00}$,$\ket{01}$,$\ket{10}$, and $\ket{11}$ refer to no error, third qubit error, second qubit error, and first qubit error, respectively.

This circuit is designed based on two stabilizer generators i.e., $S_1=Z_1\otimes Z_2$ and $S_2=Z_1 \otimes Z_3$ in the computational basis. According to the stabilizer formalism, these stabilizer generators return the eigenvalues of a quantum state. Since the eigenvalue corresponding to $\ket{0}$ and $\ket{1}$ in the computational basis is +1 and -1, respectively, $Z\otimes Z$ acts on two qubits and compares them. The generators yield +1 if two compared qubits are identical, otherwise, they return -1. The eigenvalues of these stabilizers are different for four quantum states including error-free state and quantum states with single-bit-flip error. This means that any bit-flip error converts code space to orthogonal sub-spaces. The eigenvalue of these stabilizers is equal to $(1,1)$ for the error-free case or $I\ket{\psi}$, called code-word, while each error pattern changes the eigenvalues of these stabilizers. The eigenvalue of these stabilizers for the bit-flip error on $q_0$, denoted by $X_1\ket{\psi}$, is equal to $(-1,-1)$. Likewise, the eigenvalue of these stabilizers for $X_2\ket{\psi}$ and $X_3\ket{\psi}$ is equal to $(-1,1)$ and $(-1,1)$, respectively. Having four unique eigenvalues ensures that all single bit-flip errors can be detected and corrected. It should be noted that these stabilizer generators are not unique and, for example, the error syndrome generator can be designed based on $S_1=Z_1\otimes Z_2$ and $S_2=Z_2 \otimes Z_3$ stabilizers. 

\subsection{Phase-Flip Correction}
In classical computing, information can be either zero or one at any particular instant, and all types of errors can be represented as a bit-flip error that swaps data value between these two values. There is a similar concept in quantum computing to swap values between $\ket{1}$ and $\ket{0}$ in the $Z-basis$. On the other hand, there is a new concept of error in quantum computing referring to the phase-flip swapping qubit value between $\ket{+}$ and $\ket{-}$ in the $X-basis$. Phase-flip or sign-flip is similar to the \textit{Z-gate} operation that swaps $\alpha\ket{0}+\beta\ket{1}$ and $\alpha\ket{0}-\beta\ket{1}$. Therefore, the quantum state $\ket{+}=\frac{1}{\sqrt{2}}(\ket{0}+\ket{1})$ is swapped to $\ket{-}=\frac{1}{\sqrt{2}}(\ket{0}-\ket{1})$ and vice versa. The proposed architecture is able to correct bit-flip errors caused by an \textit{X-gate} operation in the quantum channel. According to features of quantum gates, $H \circ X\circ H=Z$, by adding Hadamard gates at the output of the encoding segment and the input of the decoding segment. Therefore, the proposed architecture is able to correct phase-flip errors too. In this case the original quantum state $\ket{\psi}=\alpha\ket{0}+\beta\ket{1}$ is coded to $\ket{\psi}=\alpha\ket{+++}_{ABC}+\beta\ket{---}_{ABC}$. 

In this case, the stabilizer generators are $S_1=X_1\otimes X_2$ and $S_2=X_1 \otimes X_3$ in the $X-basis$ measurement. The eigenvalue of the code-word, $I\ket{\psi}$, is equal to $(1,1)$, while $Z_1\ket{\psi}$, $Z_2\ket{\psi}$, and $Z_3\ket{\psi}$ converts the eigenvalues to $(-1,-1)$,$(-1,1)$, and $(1,-1)$,respectively. Since the eigenvalue corresponding to $\ket{+}$ and $\ket{-}$ in the Hadamard basis, $X-basis$, is +1 and -1, respectively, $Z\otimes Z$ compares two qubits in Hadaramd basis.

\section{Evaluation and Discussion }
In this section, the effectiveness of the proposed distributed architecture is investigated in terms of fidelity obtained from simulation, complexity, and its ability to correct single errors. Some related works are also reviewed in this section. 

\subsection{Fidelity Analysis}
Fidelity is a measure to represent the closeness of two quantum states in the form of either state-vectors or density matrices, quantum unitary operations, or even quantum channels \cite{E2EFidelity}. Fidelity $(F)$ can be calculated through Equation \ref{Eq31}, which is a number between 0 and 1 $(0 \leq F \leq 1)$.
\begin{equation}\label{Eq31}
F(\rho_D,\ket{\psi}_C)=\bra{\psi_C}\rho_D\ket{\psi_C}=|\braket{\psi_D}{\psi_C}|^2
\end{equation}
where $\rho_D$  and $\ket{\psi}_C$ respectively represent the output of the proposed distributed structure and the output of the centralized structure in this context. If $F$ is equal to zero, it means that the proposed distributed structure is orthogonal to the desired centralized architecture, while If $F$ is equal to one it means that the distributed structure is the same as the centralized architecture. The fidelity assessment is conducted through the Qiskit library (version 1.1 ) on Python, running on an Intel Core (TM) i7-5500U-2.4 GHz processor and 8 GB RAM for three different structures, i.e., encoder, decoder, and quantum channel. According to the results of the evaluations, fidelity of the input quantum state and output quantum state of the proposed structures is 1.0, which proves the correctness of the proposed distributed architecture in the correction of bit-flip and phase-flip.

\subsection{Complexity of the Proposed Architecture}
In general, number of qubits, number of ancilla qubits, number of entangled qubits, number of unitary operations, and number of measurements are common performance evaluation metrics that can be utilized to compare quantum designs. The hardware complexity of the proposed distributed encoding and decoding structures are summarized in Table \ref{Table1}.
\begin{table}[htbp]
\caption{Hardware complexity of the proposed distributed structure}
\begin{adjustbox}{width=\columnwidth,center}
\begin{tabular}{cccccc}
 \hline
 \textbf{Component} & \centering\textbf{Computing } & \centering\textbf{Communication } & \textbf{Gates} & \textbf{Measurement} & \textbf{EPR } \\ 
  \textbf{} & \centering\textbf{ Qubits} & \centering\textbf{ Qubits} & \textbf{Gates} & \textbf{} & \textbf{Pairs} \\
  [0.6ex] 
 \hline\hline
 \textbf{Encoder} & \centering 3 & \centering 4 & \centering 10 & \centering 4 & \quad2\\ 
 \textbf{Decoder} & \centering 3 & \centering 4 & \centering 17 & \centering 4 &\quad4\\ [1ex] 
 \hline
\end{tabular}
\end{adjustbox}
\label{Table1}
\end{table}

In a monolithic system, single bit-flip and single phase-flip correction architectures can be implemented with five and eleven quantum gates, respectively. Although the number of quantum gates is increased in the proposed distributed architecture, the number of qubits, as the paramount limitation of quantum computing, required per QPU is diminished. On top of that, to the best of our knowledge, the proposed architecture is the first endeavor toward distributed quantum error correction and such techniques will play an essential role in distributed quantum computing.   

\subsection{Correctness Analysis of the Proposed Architecture}
In general, quantum errors can be modeled by a unitary transformation $U$ as Equation \ref{Eq32}, as a linear combination of bit-flip, phase-flip, and both. 
\begin{equation}\label{Eq32}
    U=C_0I+C_1X+C_2Y+C_3Z
\end{equation}
where, $C_0$, $C_1$, $C_2$, and $C_3$ are complex numbers, $I$, $X$, $Z$, and $Y$ refer to no error, bit-flip, phase-flip, and bit-phase-flip error, respectively. In this section, the effectiveness of the proposed distributed architecture in dealing with the bit-flip and phase-flip errors is investigated. 

In order to evaluate the bit-flip error, the quantum channel should be modeled as $U= C_1X$. The encoding segment generates a joint state including three qubits and a bit-flip can occur on each of them independently with equal probability $p$. In order to evaluate the effectiveness of the proposed distributed architecture, all possible cases of bit-flip errors are considered as shown in Table \ref{Table2}, where, $\ket{\psi}_{ABC} = \alpha\ket{000}_{ABC}+\beta\ket{111}_{ABC}$ is the code word.
\begin{table}[h]
\caption{Error pattern, quantum states, and error syndrome of the proposed architecture}
\begin{adjustbox}{width=\columnwidth,center}
\begin{tabular}{cccc}  
 \hline
 \centering\textbf{Quantum channel} & \centering\textbf{Output state of } & \centering\textbf{Probability} &\textbf{Error} \\ 
  [0.6ex] 
  \centering\textbf{(Error pattern)} & \centering\textbf{the channel $\ket{\psi'}$} & \centering\textbf{} &\textbf{Syndrome} \\ 
  [0.6ex] 
 \hline\hline
 \centering $I \otimes I \otimes I$ & \centering $\alpha\ket{000}_{ABC}+\beta\ket{111}_{ABC}$ & \centering $(1-p)^3$ &  $\ket{00}$  \\ 
  \centering $X \otimes I \otimes I$ & \centering $\alpha\ket{100}_{ABC}+\beta\ket{011}_{ABC}$ & \centering $p(1-p)^2$ &  $\ket{11}$  \\ 
   \centering $I \otimes X \otimes I$ & \centering $\alpha\ket{010}_{ABC}+\beta\ket{101}_{ABC}$ & \centering $p(1-p)^2$ &  $\ket{10}$  \\ 
  \centering $I \otimes I \otimes X$ & \centering $\alpha\ket{001}_{ABC}+\beta\ket{110}_{ABC}$ & \centering $p(1-p)^2$ &  $\ket{01}$  \\ 
  \centering $X \otimes X \otimes I$ & \centering $\alpha\ket{110}_{ABC}+\beta\ket{001}_{ABC}$ & \centering $(1-p)p^2$ &  $\ket{01}$  \\ 
  \centering $I \otimes X \otimes X$ & \centering $\alpha\ket{011}_{ABC}+\beta\ket{100}_{ABC}$ & \centering $(1-p)p^2$ &  $\ket{11}$  \\
   \centering $X \otimes I \otimes X$ & \centering $\alpha\ket{101}_{ABC}+\beta\ket{010}_{ABC}$ & \centering $(1-p)p^2$ &  $\ket{10}$  \\
    \centering $X \otimes X \otimes X$ & \centering $\alpha\ket{111}_{ABC}+\beta\ket{000}_{ABC}$ & \centering $p^3$ &  $\ket{00}$  \\ [1ex] 
 \hline
\end{tabular}
\end{adjustbox}
\label{Table2}
\end{table}

There are eight different error patterns as quantum channel features that change the quantum state of the channel’s input, $\ket{\psi}_{ABC}$. In the output quantum state of the channel, there are two different cases, i.e., the error is at most single error $(\ket{\psi}_{s})$ and the error is at least two errors $(\ket{\psi}_{t})$. The error probability of conversion $(\ket{\psi})$ to $(\ket{\psi}_{s})$ or to$(\ket{\psi}_{t})$ are equal to $(1-p)^3+3p(1-p)^2$ and $3p^2 (1-p)+p^3$, respectively. Therefore, the quantum state of the channel output can be written as Equation \ref{Eq33}.
\begin{equation}\label{Eq33}
\resizebox{.89\hsize}{!}{$
\rho_{out}=(1-p)^3+3p(1-p)^2\ket{\psi}_s\bra{\psi}_s+3p^2(1-p)+p^3\ket{\psi}_t\bra{\psi}_t$}    
\end{equation}

where,\\
$\ket{\psi}_s\bra{\psi}_s=[(I \otimes I \otimes I)+(X \otimes I \otimes I)+(I \otimes X \otimes I)+(I \otimes I \otimes X)] \otimes \ket{\psi}_{ABC}\bra{\psi}_{ABC}\\
=(\ket{000}\bra{000}+\ket{111}\bra{111})+(\ket{100}\bra{100}+\ket{011}\bra{011})+(\ket{010}\bra{010}+\ket{101}\bra{101})+(\ket{001}\bra{001}+\ket{110}\bra{110})\\
=\ket{\psi}_{ABC}\bra{\psi}_{ABC}+\sum_{i=1}^{3}X_i\ket{\psi}_{ABC}\bra{\psi}_{ABC}X_i$
and
$
\ket{\psi}_t\bra{\psi}_t=[(X \otimes X \otimes I)+(X \otimes I \otimes X)+(I \otimes X \otimes X)+(X \otimes X \otimes X)] \otimes \ket{\psi}_{ABC}\bra{\psi}_{ABC}\\
\quad=(\ket{110}\bra{110}+\ket{001}\bra{001})+(\ket{101}\bra{101}+\ket{010}\bra{010})+(\ket{011}\bra{011}+\ket{100}\bra{100}) +(\ket{111}\bra{111}+\ket{000}\bra{000})$

In general, $\ket{\psi}_s\bra{\psi}_s$ represents the states that can be corrected by the proposed distributed decoding and analysis or analysis of error syndrome, while $\ket{\psi}_t\bra{\psi}_t$ represents the states that cannot be corrected due to flipping more than one qubit. Therefore, the fidelity between $\ket{\psi}_{ABC}\bra{\psi}_{ABC}$ and $\rho_{out}$ can be analyzed through Equation \ref{Eq34}.
\begin{equation}\label{Eq34}
\begin{split}
F(\ket{\psi}_{ABC})=&F(\rho_{out},\ket{\psi}_{ABC})=\\&\bra{\psi_{ABC}}\rho_{out}\ket{\psi_{ABC}}=2p^3-3p^2+1 
\end{split}
\end{equation}

The proposed distributed architecture can detect and correct the maximum single-qubit errors if its fidelity is larger than or equal to the case when no error control technique is applied. In general, the fidelity of the system is $1-p$ if no error control techniques are applied. Therefore, $2p^3-3p^2+1$ should be larger than or equal to $1-p$ because of the error correction to be effective. This happens if $0 \leq p \leq \frac{1}{2}$. This means that the proposed distributed architecture is an effective single bit/phase-flip correction structure when the error probability is small enough, it should be noted that error syndrome is not unique to the eight possible cases that can occur. For example, two different error patterns generate error syndrome equal to $\ket{11}$, i.e., erroneous first qubit case and erroneous second and third qubits case. There is no effective mechanism to distinguish these two cases only through the error syndrome, so if error syndrome is equal to $\ket{11}$ and we are sure that the error probability is small, there is only one erroneous qubit and the first qubit should be flipped in the correction process. There is a similar argument when $\ket{\psi}=\alpha\ket{+++}_{ABC}+\beta\ket{---}_{ABC}$ is converted to other states such as $\ket{\psi}_e=\alpha\ket{++-}_{ABC}+\beta\ket{--+}_{ABC}$ or $\ket{\psi}_e=\alpha\ket{+-+}_{ABC}+\beta\ket{-+-}_{ABC}$.

The correctness of the proposed distributed architecture against phase-flip error is depicted in Table \ref{Table3}. According to the table, the initial state can be either $\ket{0}$ or $\ket{1}$ that is mapped to $\ket{+++}$ or $\ket{---}$, respectively. In general, three possible phase-flip errors can occur in each of the three qubits of the joint state, which are generated by $H \circ X \circ H=Z$. According to the captured output qubits, qubit \#0 as the disjoint state is always equal to the initial state, and ancilla qubits have returned to $\ket{0}$, as the initial state.
\begin{table*}
\caption{Phase-flip correction of the proposed distributed architecture}
\begin{center}
\begin{tabular}{ |p{0.8cm}|ccc|p{.95cm}|c|cc|}
 \hline
 Initial State & \multicolumn{3}{|c|}{Joint State} & \centering {Error Patterns} & \centering Output State & \multicolumn{2}{|c|}{\centering{Ancilla Qubits}}\\
 \hline
  &  &  &  & & &  &\\
  \multirow{3}{4em}{$\ket{0}$} & \multirow{3}{4em}{\includegraphics[width=.075\textwidth]{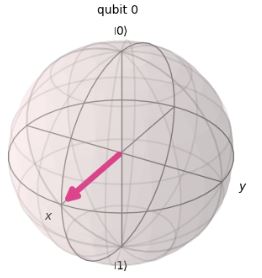}} & \multirow{3}{4em}{\includegraphics[width=.075\textwidth]{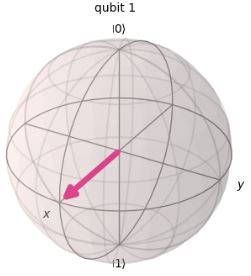}} & \multirow{3}{4em}{\centering \includegraphics[width=.075\textwidth]{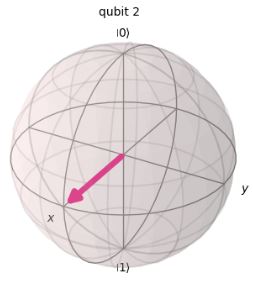} } & \centering Z(0) & \multirow{3}{4em}{\centering \includegraphics[width=.075\textwidth]{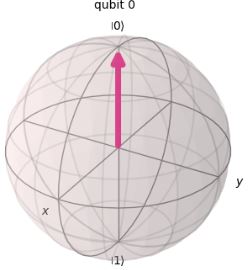}} & \multirow{3}{4em}{\centering \includegraphics[width=.075\textwidth]{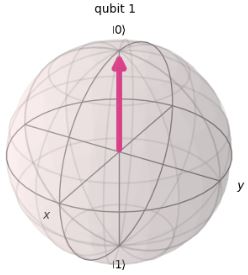}} & \multirow{3}{4em}{\centering \includegraphics[width=.075\textwidth]{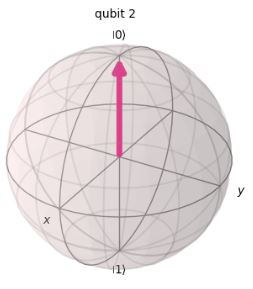}} \\[0.3em]
 & &  & & \centering Z(1) &  &  &  \\[0.3em]
 & &  & & \centering Z(2) &  &  &  \\[0.7em]
 &  &  &  & & &  &\\
 \hline
 &  &  &  & & &  &\\
 \multirow{3}{4em}{$\ket{1}$} & \multirow{3}{4em}{\centering \includegraphics[width=.075\textwidth]{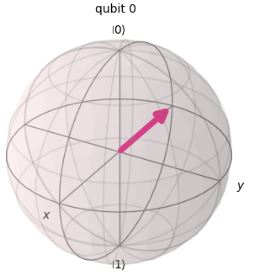}} & \multirow{3}{4em}{\centering \includegraphics[width=.075\textwidth]{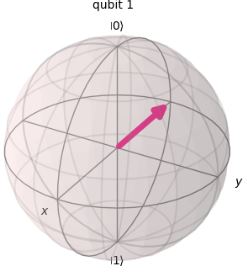}} & \multirow{3}{4em}{\centering \includegraphics[width=.075\textwidth]{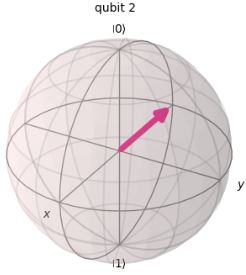}} & \centering Z(0) & \multirow{3}{4em}{\centering \includegraphics[width=.075\textwidth]{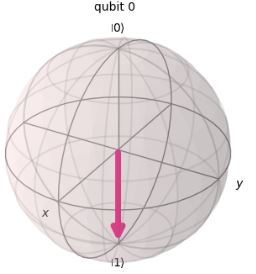}} & \multirow{3}{4em}{\centering \includegraphics[width=.075\textwidth]{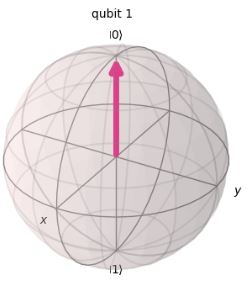}} & \multirow{3}{4em}{\centering \includegraphics[width=.075\textwidth]{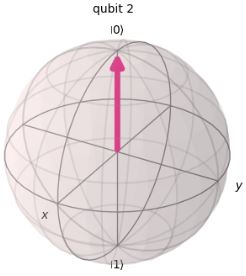}} \\[0.3em]
 & &  & & \centering Z(1) & &  &  \\[0.3em]
 & &  & & \centering Z(2) & & &  \\[0.7em]
 &  &  &  & & &  &\\
 \hline

  \multicolumn{5}{|c|}{}  & \multicolumn{3}{|c|}{Disjoint States} \\
 \hline
 \end{tabular}
\end{center}
\label{Table3}
\end{table*}

\subsection{Discussion}
The proposed distributed structure for bit/phase-flip error correction is the first work conducted at the qubit level. Therefore, there is no effective way to compare it with comparative methods in terms of evaluation metrics of distributed quantum computing. The effectiveness of the proposed structure has been analyzed in all possible combinations of bit-flip and phase-flip errors. According to the stabilizer generators, i.e., $S_1=Z_1\otimes Z_2$ and $S_2=Z_1 \otimes Z_3$ in the computational basis, and $S_1=X_1\otimes X_2$ and $S_2=X_1 \otimes X_3$ in the Hadamard basis, any single error transforms the $\ket{\psi}$ as the code-word into a distinct sub-spaces, causing a precise correction in the proposed architecture. The fidelity of the input quantum state and output quantum state has been shown to be high. The complexity in terms of gates/qubits is shown to be higher than those required if a logical qubit is implemented on a single QPU, e.g., five to eleven gates and three physical qubits, but these are still reasonable. As a future work, we will use these numbers as baseline and explore potentially better or even optimal solutions. Note that the proposed distributed architecture reduces the average number of physical qubits required per QPU to construct a logical qubit in comparison to monolithic quantum computing, which promises to increase the number of logical qubits in future quantum products.  In addition, by distributing the three physical qubits among three QPUs, the probability of correlated errors among these three physical qubits will be less likely.    

\section{Related Works}
A majority of available products on \textit{NISQ} do not leverage any known QEC codes. They rely on some classical methods such as multiple executions, applying statistical analysis, and ignoring outputs with lower occurrence rates to achieve dependable outputs, while leveraging error correction techniques increases the accuracy of outputs, enhances the service rate and efficiency of the system, and eliminates time redundancy and wastage of computing resources. 
In general, QEC techniques are classified into two categories, i.e., qubit-based and techniques quarks' feature-based techniques. Qubit-based techniques consider qubits as the basic unit of a quantum system and utilize redundancies such as logical qubits, joint states, and entanglement to protect them. Quarks' feature-based techniques attempt to exploit intrinsic quantum features of quarks such as photons, electrons, and ions for QEC. The second category of QEC utilizes the physical features of quarks such as photons and spins of electrons for more calibrations and QEC.

The first well-designed QEC code for correcting arbitrary single-qubit errors was introduced by Shor in 1995, utilizing nine physical qubits as one logical qubit. A seven-qubit code and a five-qubit code have been introduced by Steane and Laflamme, respectively, Calderbank, Shor, and Steane developed a general class of code, called CSS codes for single error correction. According to Hamaming bound, it has been proven that at least five physical qubits are required for arbitrary error correction. Due to the importance of error correction in quantum technology, several codes such as additive codes, lattice codes, and toric codes have been developed in the last two decades. Entanglement-assisted QEC codes are other types of codes that exploit the features of tightly correlated entangled particles to achieve fault-tolerant systems. Surface codes are another powerful family of quantum error-correcting codes that utilize a 2D lattice of qubits as logical qubits and are considered promising \cite{new4}.

In recent years, several innovative QECs have been proposed that require specific requirements. In the quantum secret sharing (QSS) technique, the original qubit is encoded into three sharings to retrieve the erased qubit, in which the erased qubit can be retrieved through two out of three sharings \cite{10092522}. In this technique, \textit{qutrit} is the basic unit of quantum information, which is a linear combination of three orthogonal quantum states, i.e., $\ket{0}$, $\ket{1}$, and $\ket{2}$. Saraiva and Bartlett introduced a three-spin-qubit silicon device together with a novel quantum gate to cover common errors \cite{new2}. Takeda et al. introduced silicon spin qubits-based quantum error correction due to their compatibility with mature nanofabrication technologies \cite{new1}. Furthermore, since qubit-based coding is not effective for systems that store information in bosonic systems such as photonic resonators, Jain et al., introduced quantum spherical codes for bosonic, spin, and molecular systems. They have claimed that their polytope-based cat codes are multi-mode extensions of the cat codes and can outperform previous constructions \cite{new3}. 

All previous works on QEC assumed that a logical qubit is formed using multiple physical qubits residing on a QPU. Not only the limitation of the number of physical qubits on the available QPUs further limits the number of logical qubits, but the physical qubits on the same QPU will likely suffer from correlated errors, affecting QEC performance. To the best of our knowledge, the proposed structure is the first work that considered distributed QEC.

\section{Conclusion}
Although IBM released the first-ever 1,000-qubit quantum chip in December 2023, there is still a long way to go before one has enough error-free qubits for real quantum computing applications. A new distributed bit/phase-flip error correction mechanism has been proposed in this paper, which is in line with the transition from centralized quantum computing to distributed quantum computing. Three quantum processing units collaboratively create a joint quantum state where a single bit-flip and phase-flip errors are corrected through the decoding and syndrome analysis segments. According to the syndrome analysis, error patterns of the quantum channel, and fidelity analysis, the proposed qubit-based design is able to handle all single errors efficiently. The implementation of other quantum error correction codes such as surface codes, Shor’s code, and CSS code in the distributed structure can be considered as future works of this research.

\end{document}